\documentclass[12pt]{article}

\usepackage{graphicx}

\def\theequation{\thesection.\arabic{equation}}
\newcommand{\beq}{\begin{equation}}
\newcommand{\eeq}{\end{equation}}
\newcommand{\bea}{\begin{eqnarray}}
\newcommand{\eea}{\end{eqnarray}}

\newcommand{\be}{\begin{equation}}
\newcommand{\ee}{\end{equation}}
\newcommand{\ba}{\begin{array}}
\newcommand{\ea}{\end{array}}

\newcommand{\al}{\alpha}

\newcommand{\de}{\delta}

\newcommand{\ep}{\epsilon}

\newcommand{\ta}{\tau}

\newcommand{\La}{\Lambda}

\newcommand{\rar}{\rightarrow}
\newcommand{\non}{\nonumber}
\newcommand{\ts}{\textstyle}

\def\eqs#1{(\ref{#1})}

\font\cmss=cmss10 at 11pt \font\cmsss=cmss8 at 8pt

\def\inbar{\vrule height1.5ex width.4pt depth0pt}
\def\mininbar{\vrule height.75ex width.3pt depth0pt}
\def\cc{\relax\,\hbox{$\mininbar\kern-.2em{\hbox{\rm\tiny C}}$}}
\def\IZ{\relax\ifmmode\mathchoice
{\hbox{\cmss Z\kern-.4em Z}}{\hbox{\cmss Z\kern-.4em Z}}
{\lower.4pt\hbox{\cmsss Z\kern-.4em Z}}
{\lower1.2pt\hbox{\cmsss Z\kern-.4em Z}}\else{\cmss Z\kern-.4em Z}\fi}
\def\IC{\relax\,\hbox{$\inbar\kern-.3em{\rm C}$}}
\def\IR{\relax{\rm I\kern-.18em R}}

\newcommand{\SU}{\mathrm{SU}}

\newcommand{\U}{\mathrm{U}}

\newcommand{\PP}{\mathrm{I}\kern -2pt \mathrm{P}}
\newcommand{\R}{\mathrm{I}\kern -2.5pt \mathrm{R}}
\newcommand{\Z}{\mathsf{Z}\kern -5pt \mathsf{Z}}
\newcommand{\1}{1\kern -3pt \mathrm{l}}

\newcommand{\tr}{{\rm tr}}

\newcommand{\pa}{\partial}

\newcommand{\cO}{\mathcal{O}}
\newcommand{\cN}{\mathcal{N}}
\newcommand{\cF}{\mathcal{F}}

\newcommand{\D}{{\rm d}}
\newcommand{\e}{{\rm e}}

\def\half{ {\textstyle\frac{1}{2}} }

\def\eps{ \epsilon }
\def\Sig{\Sigma}

\def\minus{ - }
\def\minusp{ - }
\def\plus{ + }
\def\muteplus{ }
\def\hLa{\hat{\Lambda}}
\def\pre{ \cF(a) }
\def\laSW{ \lambda_{SW} }
\def\gs{ g_{s} }
\def\free{ F_0(e, S) }
\def\freen{ F^{(n)}_0(e, S) }
\def\freethree{ F^{(3)}_0(e, S) }
\def\Weff{ W_{\rm eff}}
\def\tauo{\tau_0} 
\def\gpi{ \gamma_{p,i} }
\def\eij{  e_{ij} }
\def\eik{  e_{ik} }
\def\ejk{  e_{jk} }
\def\eji{  e_{ji} }
\def\eil{  e_{i\ell} }
\def\ekl{  e_{k\ell} }
\def\aij{  a_{ij} }
\def\aik{  a_{ik} }
\def\ajk{  a_{jk} }
\def\aji{  a_{ji} }
\def\ail{  a_{i\ell} }

\def\tLa{\tilde{\La}}
\def\tSig{\tilde{\Sigma}}
\def\te{ \tilde{e}}
\def\ta{ \tilde{a}}
\def\tA{ \tilde{A}}
\def\tlaSW{ \tilde{\lambda}_{SW} }
\def\tW{ \tilde{W} }
\def\tZ{ \tilde{Z} }
\def\tS{ \tilde{S} }
\def\tu{ \tilde{u} }
\def\tWeff{ \tilde{W}_{\rm eff} }
\def\tfree{ \tilde{F}_0 (e, S, \eps) }
\def\delF{\delta F}
\def\delS{\delta S}

\def\sumN{ \sum_{i=1}^{N} }
\def\sumK{ \sum_{j=1}^{K} }
\def\prodN{ \prod_{i=1}^{N} }
\def\prodkN{ \prod_{k=1}^{N} }
\def\prodK{ \prod_{j=1}^{K} }
\def\sumk{ \sum_{k\neq i} }
\def\suml{ \sum_{\ell \neq i,k} }
\def\planar{ \bigg|_{\rm planar} }
\def\vev#1{ \langle {#1} \rangle }
\def\tadpole{ \vev{\tr(\Psi_{ii})}}
\def\vevS{ \bigg|_{\vev{S}}  }
\def\argu{ \left[ \pi (j-\half)/K  \right] }

\begin{document}

\begin{flushright}
BRX-TH-506\\
BOW-PH-126\\
{\tt hep-th/0211123}
\end{flushright}
\vspace{.3in}
\setcounter{footnote}{0}
\stepcounter{table}

\begin{center}

{\Large{\bf\sf The {\large $\cN=2$ $\U(N)$} gauge 
theory prepotential and periods 
\\ from a perturbative matrix model calculation}}

\vspace{.2in}

Stephen G. Naculich\footnote{Research
supported in part by the NSF under grant PHY-0140281.}$^{,a}$,
Howard J. Schnitzer\footnote{Research
supported in part by the DOE under grant DE--FG02--92ER40706.}$^{,b}$,
and Niclas Wyllard\footnote{Research 
supported by the DOE under grant DE--FG02--92ER40706.\\
{\tt \phantom{aaa} naculich@bowdoin.edu;
schnitzer,wyllard@brandeis.edu}\\}$^{,b}$

\vspace*{0.3in}

$^{a}${\em Department of Physics\\
Bowdoin College, Brunswick, ME 04011}

\vspace{.2in}

$^{b}${\em Martin Fisher School of Physics\\
Brandeis University, Waltham, MA 02454}

\end{center}

\vskip 5mm

\begin{abstract}
We perform a completely perturbative matrix model 
calculation of the physical low-energy quantities of
the $\cN=2$ $\U(N)$ gauge theory.  
Within the matrix model framework we propose a 
perturbative definition of the periods $a_i$ in terms of certain 
tadpole diagrams, and check our conjecture 
up to first order in the gauge theory instanton expansion. 
The prescription does not require knowledge of 
the Seiberg-Witten differential or curve. 
We also compute the $\cN=2$ prepotential $\cF(a)$ 
perturbatively up to the first-instanton level,
finding agreement with the known result.

\end{abstract}

\section{Introduction}
\setcounter{equation}{0}

Dijkgraaf and Vafa, drawing on earlier developments
\cite{Bershadsky:1993}--\cite{Cachazo:2002}, 
have uncovered the surprising result that 
non-perturbative effective superpotentials for certain 
$d=4$ $\cN=1$ supersymmetric gauge theories 
can be obtained by calculating planar diagrams 
in a related gauged matrix 
model~\cite{Dijkgraaf:2002a}--\cite{Dijkgraaf:2002d}.
In particular, the $d$-instanton contribution to the 
effective superpotential can be obtained from the calculation of 
$(d{+}1)$-loop planar diagrams in an associated matrix model.
The simplest example is the $\cN=1$  SU$(N)$ gauge theory with 
an adjoint chiral superfield $\phi$ and tree-level superpotential $W(\phi)$,  
for which the instanton corrections can be obtained from the calculation of 
the planar loop diagrams in a hermitian matrix model. This 
statement has recently been proven~\cite{Dijkgraaf:2002e}. 
Further work along these lines has been
presented in refs.~\cite{multifarious}.

The new approach can also be used to study 
$d=4$ $\cN=2$ supersymmetric gauge 
theories,
by using $W(\phi)$ to freeze the moduli at an
arbitrary point on the Coulomb branch of the $\cN=2$ theory, 
thereby breaking $\cN=2$ to $\cN=1$,
and then turning off $W(\phi)$ at the end of the calculation 
to restore $\cN=2$ supersymmetry~\cite{Cachazo:2002}--\cite{Dijkgraaf:2002c}.
The crucial feature that makes this work is that certain quantities 
are independent of the parameter that goes to zero in the limit when 
$\cN=2$ supersymmetry is restored, and 
can thus be calculated for finite values of the parameter.

Even when the matrix model cannot be completely solved,
a perturbative diagrammatic expansion of the matrix model 
can still be used to obtain non-perturbative 
information about the $\cN=2$ gauge theory.
In ref.~\cite{Dijkgraaf:2002d}, 
the effective gauge coupling matrix $\tau_{ij}$ 
of the unbroken $\U(1){\times} \U(1)$ gauge group 
at an arbitrary point on the Coulomb branch of the $\cN=2$ U(2) gauge
theory was computed, as a function of the classical modulus,
to several orders in the instanton expansion\footnote{For 
this case, the exact all-orders result can be obtained from 
the known large-$M$ two-cut solution of the matrix 
model~\cite{Dijkgraaf:2002a,Dijkgraaf:2002c, Dijkgraaf:2002d}.}.

In this paper, we extend this result to the $\cN=2$ U($N$) gauge 
theory, computing the matrix of effective gauge couplings $\tau_{ij}$
of the unbroken U$(1)^{N}$ gauge group 
as a function of the classical moduli, which we denote by $e_i$.
To explicitly obtain the full low-energy physical content of the model, 
however,
one also needs to determine the relation between 
the periods $a_i$ and the classical moduli $e_i$.
We argue that $a_i$ can be determined by computing 
tadpole diagrams in perturbative matrix theory, 
and verify that this prescription yields the 
correct results for pure U($N$) gauge theory up through one-instanton.
Knowing the connection between $a_i$ and $e_i$ 
enables us to re-express $\tau_{ij}$ as a function of $a_i$.
This then allows the relations
$\tau_{ij} (a) = {\pa a_{D,i} / \pa a_j} = {\pa^2 \cF(a) / \pa a_i \pa a_j} $
to be integrated.
Thus, we demonstrate that exact nonperturbative quantities 
in  low-energy $\cN=2$ supersymmetric theories,
namely, the prepotential $\cF(a)$ and 
the masses of BPS states $|n a + m a_D|$,
can be computed from a diagrammatic expansion of the matrix model,
even in cases when an exact solution of the matrix model is not known.

Solving for the gauge coupling matrix, 
prepotential and BPS mass spectrum perturbatively, 
without using the exact solution of the matrix model,
is equivalent to deriving these results without knowledge 
of the Seiberg-Witten curve or differential
(although they are known in the particular case we study).
Thus the techniques developed here and in 
refs.~\cite{Dijkgraaf:2002a}-\cite{Dijkgraaf:2002d} 
could be used to obtain non-perturbative information
about $\cN=2$ supersymmetric gauge theories for which the Seiberg-Witten
curve is not known.

In sec.~2, we review the Seiberg-Witten approach to
the calculation of the prepotential, periods, 
and gauge couplings in $\cN=2$ gauge theories.
In sec.~3, we describe the matrix model approach to the 
calculation of the gauge coupling matrix $\tau_{ij}$, 
and in sec.~4 we carry out the calculation of $\tau_{ij}$ 
to one-instanton order for the $\cN=2$ U($N$) gauge theory.
In sec.~5, we present our proposal for computing $a_i$ 
in the perturbative matrix model,
and in sec.~6 we compute the relation between $a_i$ and $e_i$
up to one-instanton for U($N$).  
Using this result together with the results of sec.~4,
we compute the $\cN=2$ prepotential $\cF(a)$ to one-instanton level. 
Finally, in sec.~7, we calculate the gauge theory invariants
$\vev{\tr(\phi^n)}$ perturbatively in the matrix model,
finding agreement with known results. 
In an appendix, we present an alternative method of computing
the relation between $a_i$ and $e_i$ 
using the relation between the Seiberg-Witten differential 
and the density of gauge theory eigenvalues in the large-$N$ 
limit \cite{Dijkgraaf:2002d}.

\section{Seiberg-Witten approach to $\cN=2$ gauge theories}
\setcounter{equation}{0}

The Seiberg-Witten approach to $\cN=2$ supersymmetric gauge 
theory \cite{Seiberg}
involves identifying a complex curve $\Sig$ and a meromorphic differential 
$\laSW$ on this curve.
For pure SU($N$) gauge theory\footnote{
This is the nontrivial piece of the $\U(N)$ gauge theory (in later 
sections we focus on the $\U(N)$ theory).} 
the curve is given by a genus $N-1$ hyperelliptic Riemann 
surface \cite{SUN}--\cite{Douglas:1995}  
\beq
\label{Sigma}
\Sig:  \hskip.1in  y^2 = P_N (x)^2 - 4 \La^{2N}\,;  \hskip.3in
P_N(x) = \sum_{\ell=0}^N s_{N-\ell}(e) x^\ell = \prod_{i=1}^N (x-e_i)\,; 
\quad\quad  
\sumN e_i = 0\,,
\eeq
corresponding to a generic point on the Coulomb branch of the
moduli space of vacua, where the gauge symmetry is broken to U$(1)^{N-1}$.
In the equation above, $s_m(e)$ is the elementary symmetric polynomial  
\be \label{spoly}
s_m(e) = (-1)^m \sum_{i_1 < i_2 < \cdots < i_m} 
e_{i_1} e_{i_2} \cdots e_{i_m}\,, \qquad\qquad s_0 = 1\,.
\ee
Next, one chooses a canonical homology 
basis of $\Sigma$, $\{A_i, B_i\}$, $i=1, \cdots, N-1$,
in terms of which
\beq
\label{periods}
a_i = {1\over 2\pi i} \oint_{A_i} \laSW\,, \hskip.3in
a_{D,i} ={1\over 2\pi i}  \oint_{B_i} \laSW\,, \hskip.3in
\laSW = x  {\D y \over y} 
= { x P_N^\prime (x) \D x \over \sqrt{P_N (x)^2 - 4 \La^{2N}}}\,.
\eeq
We will choose $A_i$, $i=1, \cdots, N-1$  
to be the contour that remains on one sheet 
of the two-sheeted Riemann surface 
and encircles the branch cut emanating from $e_i$ \cite{D'Hoker:1996}.
$A_N$ and $a_n$ are defined similarly. 
However, $A_N$ is not an independent cycle, being equivalent 
to $ - ~\sum_{i=1}^{N-1} A_i$,
and one can show that 
$\sum_{i=1}^N a_i =\sum_{i=1}^N e_i$ by deforming the contour
and evaluating the residue of $\laSW$ at infinity.

The $A_i$-period integral may be inverted to write $e_i$ in terms of $a_i$,
allowing one to express $a_{D,i}$ as a function of $a_i$.
Then, since 
$\partial a_{D,i}/\partial a_j = \partial a_{D,j}/\partial a_i $,
one may write 
\beq
a_{D,i} = {\partial \pre \over \partial a_i}, \hskip.5in
\pre = \cF_{\rm pert} (a, \log \La) 
+ \sum_{d=1}^\infty \La^{2Nd} \cF^{(d)} (a)\,,
\eeq
thus defining the $\cN=2$ prepotential $\pre$,
which can be written as a sum of perturbative and
instanton contributions.
The masses of the BPS states of the theory
can be expressed as  $| n a + m a_{D} |$, for integers $n$, $m$.
Finally,
\beq
\label{gaugetau}
\tau_{ij} (a) = { \partial^2 \pre \over \partial a_i \partial a_j}\,,
\eeq 
yields the period matrix of $\Sig$,
identified with the gauge couplings 
of the U$(1)^{N-1}$ factors of the unbroken gauge theory.

\section{Matrix model approach to $\cN=2$ gauge theories}
\setcounter{equation}{0}

In this section we describe the matrix model approach to 
$\cN=2$ supersymmetric $\U(N)$ gauge theory. The first step is 
to break $\cN=2$ to $\cN=1$ by the addition of a tree-level
superpotential $W(\phi)$ to the gauge theory. 
This superpotential is identified with the potential of a chiral 
matrix model~\cite{Dijkgraaf:2002a}-\cite{Dijkgraaf:2002d}.
The matrix model thus has the partition function 
\cite{Dijkgraaf:2002a}-\cite{Dijkgraaf:2002d}
\beq
\label{partition}
Z = {1\over \mathrm{vol}(G)} 
\int \D\Phi \exp \left( \minus {W(\Phi)\over \gs} \right) \,,
\eeq
where the integral is over $M {\times} M$ matrices $\Phi$ 
(which can be taken to be hermitian), 
$\gs$ is a parameter that later will be taken to zero
as $M \to \infty$,
and $G$ is the unbroken matrix model gauge group.
One chooses a superpotential $W(\Phi)$ that freezes 
the moduli to a generic point on the Coulomb branch
of the $\cN=2$ theory:
\beq \label{W}
W(\Phi) = \al \sum_{\ell=0}^{N} {s_{N-\ell}(e)\over \ell{+}1} 
\tr (\Phi^{\ell+1} )
\quad \Rightarrow \quad 
W^\prime (x)  = \al \prodN (x-e_i)\,,
\eeq
where $s_m(e)$ was defined in eq.~(\ref{spoly}), 
and $\al$ is a parameter that will be taken to zero 
at the end of the calculation, 
restoring $\cN\!=2$ supersymmetry.
The matrix integral \eqs{partition} is evaluated perturbatively 
about the extremum 
\beq
\label{Phinought}
\Phi_0 = \pmatrix{ e_1 \1_{M_1}& 0& \cdots& 0 \cr
                   0& e_2 \1_{M_2}& \cdots& 0 \cr
                   \vdots& \vdots& \ddots& \vdots \cr
                   0& 0& \cdots&  e_N \1_{M_N} } \,,
\quad {\rm where~} \quad \sumN M_i = M\,,
\eeq
which breaks the $\U(M)$ symmetry to $G = \prod_{i=1}^N  U(M_i)$. 
(This is the matrix model analog of the gauge theory breaking 
$\U(N) \rar \U(1)^N$. Note that in the matrix model $M_i \gg 1$ 
for all $i$.)

Using the standard double-line notation, the connected diagrams of 
the perturbative expansion of $Z$ may be organized in an 
expansion characterized by the genus $g$ of the surface
in which the diagram is  
embedded \cite{tHooft:1974}
\beq
Z =\exp \left(   \sum_{g \ge 0} \gs^{2g-2} F_{g} (e, S) \right)
\quad {\rm where~} \quad S_i \equiv \gs M_i\,.
\eeq
Evaluating the matrix integral in the $M_i \to \infty$, $\gs \to 0$  limit,
with $S_i$ held fixed,
is equivalent to retaining only the planar (genus $g=0$) diagrams.
Thus 
\beq
\free  = \gs^2 \log  Z \planar
\eeq
corresponds to the connected planar diagrams of the matrix theory. 

To relate this to the $\cN=2$ U($N$) gauge theory 
broken to $\prod_i \U(N_i)$, 
one introduces \cite{Cachazo:2001}--\cite{Dijkgraaf:2002c}
\beq
\label{Weffdef}
\Weff (e, S) 
= \minus \sum_i  N_i {\partial \free \over \partial S_i}
  \plus 2 \pi i \tau_0 \sum_i  S_i
\eeq
where $\tau_0 = \tau(\La_0)$ is the gauge coupling 
of the U$(N)$ theory at some scale $\La_0$.
In this paper, we are interested in
breaking $\U(N)$ to $\U(1)^N$, so $N_i=1$ for all $i$, 
and $i$ runs from 1 to $N$.
The effective superpotential is extremized with respect to $S_i$ 
to obtain $\vev{S_i}$:
\beq
{\partial \Weff (e, S)  \over \partial S_i} \bigg|_{S_j = \vev{S_j} } =0\,.
\eeq
Finally, 
\beq
\label{matrixtau}
\tau_{ij} (e) = {1\over 2\pi i} 
{\partial^2 \free \over \partial S_i \partial S_j}
\bigg|_{S_i = \vev{S_i} } 
\eeq
yields the couplings of the unbroken U$(1)^N$ factors of the gauge theory,
as a function of $e_i$.  
At the end of the matrix model calculation, 
one must take $\al \to 0$ to restore $\cN=2$ supersymmetry, 
but as will be seen,
$\tau_{ij}$ is independent of $\al$, and can thus be calculated 
for any value of $\al$.

In the next section, we will explicitly carry out the procedure 
outlined above for the pure $\cN=2$ U($N$) gauge theory.

Despite the superficial similarity of 
eqs.~\eqs{gaugetau} and \eqs{matrixtau},
the $\cN=2$ gauge theory prepotential 
$\pre$ and the free energy $\free$ 
of the large $M_i$ matrix model are conceptually distinct.
$\pre$ is a function of the periods $a_i$ of the Seiberg-Witten
differential,
whereas $\free$ is a function of the $e_i$'s as well as 
the auxiliary parameters $S_i$ 
(which can understood as SU$(K)$ glueball superfields 
in the related $\U(NK) \to \U(K)^N$ theory \cite{Cachazo:2001}).
Although both \eqs{gaugetau} and \eqs{matrixtau} correspond to 
the same quantity (the period matrix of $\Sig$),
they are expressed in terms of different parameters ($a_i$ vs. $e_i$) 
on the moduli space.

If we are to use the matrix model result \eqs{matrixtau} to determine
the  $\cN=2$ prepotential $\pre$, 
we must first express $\tau_{ij}$ in terms of $a_i$.
Although the relationship between $a_i$ and $e_i$ is straightforwardly 
obtained \cite{D'Hoker:1996} in the Seiberg-Witten approach
from the $A_i$-period integral \eqs{periods},
we wish to derive this relationship from within the matrix model, without
referring to the Seiberg-Witten curve or differential.
After explicitly calculating $\tau_{ij}$ for U$(N)$ in the next section,
we will turn to a perturbative matrix model calculation
of $a_i$ for that same model in section 5.

\section{Calculation of $\tau_{ij}$ for $\U(N)$ using the matrix model}
\setcounter{equation}{0}

In this section, we will evaluate the planar free energy $\free$,
defined via
\beq
\label{Z}
\exp \left(   {1\over \gs^2} \free \right)
= {1\over \mathrm{vol}(G)} 
\int \D\Phi \exp \left( \minus \, {W(\Phi)\over \gs} \right) \planar
\eeq
to cubic order in $S_i$.
This will enable us to calculate the gauge coupling matrix $\tau_{ij}$
for $\cN=2$ U($N$) gauge theory to one-instanton accuracy.

As described in the previous section, 
we expand $\Phi$ about the following extremum of $W(\Phi)$,
\bea
\label{expand}
\Phi = \Phi_0 + \Psi = \pmatrix{ e_1 \1_{M_1}& 0& \cdots& 0 \cr
                                 0& e_2 \1_{M_2}& \cdots& 0 \cr
                                 \vdots& \vdots& \ddots& \vdots \cr
                                 0& 0& \cdots&  e_N \1_{M_N} }
                     + \pmatrix{ \Psi_{11}& \Psi_{12}& \cdots& \Psi_{1N}  \cr
                                 \Psi_{21}& \Psi_{22}& \cdots& \Psi_{2N}  \cr
                                 \vdots& \vdots& \ddots& \vdots  \cr
                                 \Psi_{N1}& \Psi_{N2} & \cdots&  \Psi_{NN} }
\eea
where $\Psi_{ij}$ is an $M_i \,{\times} M_j$ matrix.
This choice breaks $\U(M) \to G = \prod_{i=1}^N  \U(M_i)$.

Expanding $W(\Phi)$ to quadratic order in $\Psi$, we obtain 
\bea
W(\Phi) &=& \sumN   M_i W(e_i) +  
\half \al \sumN \left(\sum_{\ell=0}^N \ell s_{N-\ell} e_i^{\ell-1} \right)
\tr (\Psi_{ii}^2 ) \non\\
&+& \half \al \sumN \sum_{j\neq i} 
\left(\sum_{\ell=1}^N s_{N-\ell} 
\sum_{m=0}^{\ell-1} e_i^m e_j^{\ell-m-1} \right)
\tr (\Psi_{ij} \Psi_{ji} ) + \cO (\Psi^3) 
\eea
It can be shown that 
\bea
&& \sum_{\ell=0}^N \ell s_{N-\ell} e_i^{\ell-1} 
= \left[ {\pa \over \pa x} \prodkN (x-e_k)\right] \bigg|_{x=e_i} 
= \prod_{k\neq i} (e_i - e_k) \,,\non\\
&& \sum_{\ell=1}^N s_{N-\ell} \sum_{m=0}^{\ell-1} e_i^m e_j^{\ell-m-1} 
= 0\,,
\eea
which implies that the coefficient of 
$\tr (\Psi_{ij} \Psi_{ji})$ vanishes when $i\neq j$. Hence   
the off-diagonal matrices $\Psi_{ij}$ are zero modes, and correspond
to pure gauge degrees of freedom. These zero modes parametrize the coset 
$\U(M)/G = \U(\sum_i M_i)/[\U(M_1)\times \cdots \times \U(M_N)]$. 
Following ref.~\cite{Dijkgraaf:2002d}, 
we will fix the gauge $\Psi_{ij}=0$ ($i\neq j$) and introduce 
Grassmann-odd ghost matrices $B$ and $C$ 
with the action 
\beq
\tr \left(  B [\Phi, C] \right) = 
 \sumN \sum_{j \neq i} (e_i - e_j) \tr (B_{ji} C_{ij})
+\sumN \sum_{j \neq i} 
\tr (B_{ji} \Psi_{ii} C_{ij} - B_{ji} C_{ij} \Psi_{jj})\,.
\eeq
Thus the planar free energy is given in terms of the gauge-fixed integral
\beq
\label{freeenergy}
\exp \left(   {1\over \gs^2} \free \right)
 ={1\over \mathrm{vol}(G)}  
\exp \left(\minus {1\over \gs} \sumN   M_i W(e_i) \right)
 \int \D\Psi_{ii}\, \D B_{ij}\, \D C_{ij} 
\e^{I_{\rm quad} + I_{\rm int}}  \planar
\eeq
where the quadratic part of the action is
\beq \label{Iprop}
I_{\rm quad}  =  \minus \half  {\al\over\gs} \sumN  R_i \tr(\Psi_{ii}^2) 
\minusp \sumN \sum_{j\neq i} \eij \tr (B_{ji} C_{ij} )\,,
\qquad
R_i  =  \prod_{j\neq i} \eij\,,
\qquad
\eij = e_i - e_j
\eeq
and the interaction terms are
(after implementing the gauge choice $\Psi_{ij}=0$)
\beq 
\label{Iint}
I_{\rm int} 
=  \minus {\al\over\gs} \sumN \sum_{p=3}^N  {\gpi \over p} \tr(\Psi_{ii}^p)
\minusp  \sumN \sum_{j \neq i} 
\tr (B_{ji} \Psi_{ii} C_{ij} - B_{ji} C_{ij} \Psi_{jj})\,.
\eeq
Here
\beq
\gpi = {1\over (p-1)! }
\left[ \left( \pa \over \pa x\right)^{p-1} \prodkN  (x-e_k) \right] 
\Bigg|_{x=e_i}
\eeq
and, in particular, we will need
\beq
\gamma_{3,i} = R_i \sumk {1\over\eik}\,, \qquad
\gamma_{4,i} = \half R_i \sumk \suml {1\over\eik \eil}\,.
\eeq
The $\Psi_{ii}$ and ghost propagators can be derived
from eq.~\eqs{Iprop} and the vertices from eq.~\eqs{Iint}.
Each ghost loop will acquire an additional factor of 
$(-2)$~\cite{Dijkgraaf:2002d}. 

For large $M_i$, the volume prefactor in eq.~\eqs{freeenergy}
becomes \cite{Ooguri:2002} 
\beq
{1\over \mathrm{vol}(G)} 
= \exp \left( \half \sumN {M_i^2 } \log M_i \right)\,.
\eeq
The integral of the quadratic action $I_{\mathrm{quad}}$ may 
be evaluated to give 
\beq
\prodN \left(  \gs \over \alpha R_i \right)^{{1\over 2}  M_i^2}
\prodN \prod_{j\neq i} \left( \eij  \right)^{M_i M_j}
\eeq
up to some multiplicative factors.
Thus, setting $S_i = \gs M_i$,  
the matrix integral \eqs{freeenergy}
yields the planar free energy
(up to a quadratic monomial in the $S_i$'s)
\beq
\label{freeenergytwo}
\free = 
\minus \sumN S_i W(e_i) 
+ \half \sumN S_i^2 \log \left( S_i\over \alpha R_i \hLa^2 \right) 
+ \sumN \sum_{j\neq i}  S_i S_j \log\left(\eij \over \hLa \right) + 
\sum_{n \ge 3} \freen
\eeq
where $\freen$ is an $n$th order polynomial in $S_i$
arising from planar loop diagrams built from the interaction
vertices \cite{Dijkgraaf:2002c}.
We have included in eq.~\eqs{freeenergytwo}
a contribution $ - \left( \sumN S_i \right)^2 \log \hLa$
that reflects the ambiguity in the cut-off of the full $\U(M)$
gauge group \cite{Dijkgraaf:2002d}.
As we will see below, the first three terms 
in eq.~\eqs{freeenergytwo} are already sufficient to give the complete
perturbative (from the gauge theory perspective) 
contribution to $\tau_{ij}$ 

To obtain $\tau_{ij}$ to one-instanton accuracy in the gauge theory,
we need to evaluate the contribution to $\free$ cubic in $S_i$. 
The Feynman diagrams that contribute at this order are depicted 
in fig.~1.

\bigskip

\begin{figure}[h]
\begin{center}
 \includegraphics{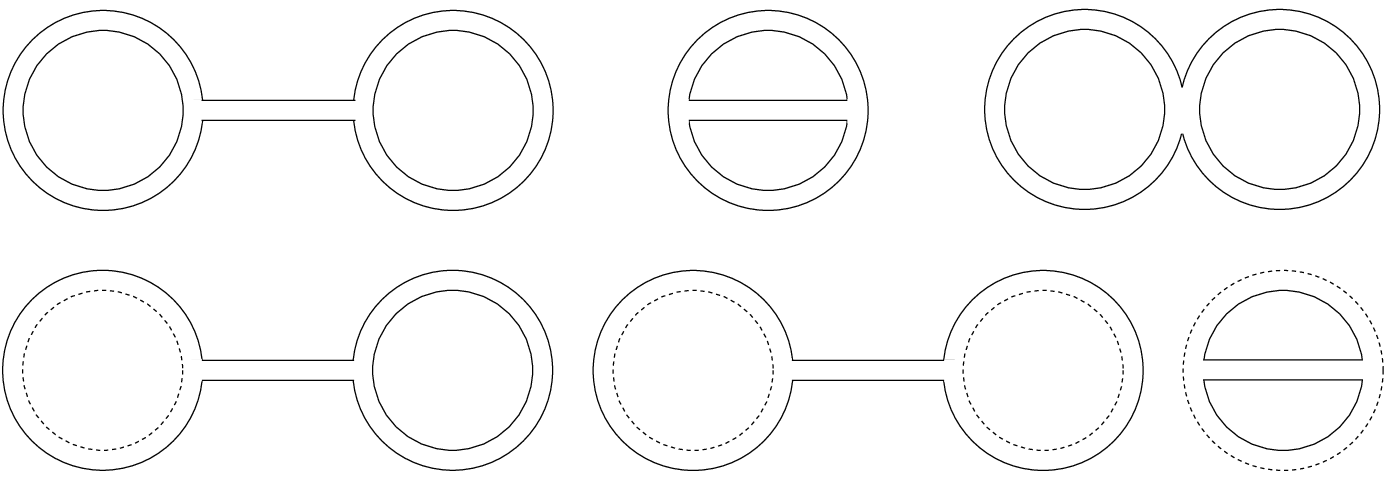}  \\[-1cm]
{\small {\bf Figure 1:} Diagrams contributing to $\free$ at $\cO(S^3)$.
Solid double lines correspond to $\Psi_{ii}$ propagators;
solid plus dashed double lines correspond to ghost propagators.}
\end{center}
\end{figure}

The six diagrams in fig. 1 give
\bea
\label{freeloop}
\alpha \freethree &=& 
({\ts \frac{1}{2} } + {\ts \frac{1}{6} } ) 
\sum_i  \frac{S_i^3}{R_i} \left( \sumk {1 \over \eik} \right)^2
\minusp {\ts \frac{1}{4} }
\sum_i \frac{S_i^3}{R_i} \sumk \suml {1 \over \eik \eil} \non\\ 
&& 
- 2 \sum_i \sumk \frac{S_i^2 S_k}{R_i \eik} \sum_{\ell \neq i} \frac{1}{\eil}  
+ 2 \sum_i \sumk \sum_{\ell \neq i} \frac{S_iS_kS_\ell}{R_i \eik \eil}
-   \sum_i \sumk  \frac{S_i^2 S_k}{R_i \eik^2}\,.
\eea
Using eq.~\eqs{freeenergytwo} and \eqs{freeloop} in eq.~\eqs{Weffdef}, 
we obtain
\bea
\label{effective}
\Weff &=&  
\sum_i W(e_i) 
\minusp \sum_i  S_i \log \left( S_i \over \alpha R_i \hLa^2 \right) 
\minusp 2 \sum_i \sumk  S_k \log \left( \eik\over \hLa \right)    \non\\
&& 
\minusp {1\over \al} \bigg[ 
\minus {\ts \frac{3}{4}} \sum_i \sumk \suml \frac{S_i^2}{R_i \eik \eil} 
+ 2 \sum_i \sumk \suml \frac{S_kS_\ell}{R_i \eik\eil} 
- \sum_i \sumk \frac{S_i^2}{R_i\eik^2}  \non\\
&&
- 2 \sum_i \sumk \frac{S_i S_k}{R_i \eik^2 } 
+ 2\sum_i \sumk \frac{S_i^2}{R_k \eik^2}  \bigg]
\plus \left( 2\pi i \tauo + {\rm const}\right) \sum_i S_i \,.
\eea
Extremizing this with respect to $S_i$ yields the equation
\bea
\label{extreme}
0 &=& \log \left( S_i R_i \over \alpha \hLa^{2N} \right)  
+ {1\over \al} \Bigg[ 
\minus  {3\over 2}  \frac{S_i}{R_i} \sumk \suml {1\over \eik\eil}
- 4 \sumk \suml \frac{S_\ell}{R_k \eik \ekl} 
- 2 \frac{S_i}{R_i} \sumk \frac{1}{\eik^2}  \non\\
&&
- \frac{2}{R_i} \sumk \frac{S_k}{\eik^2}
- 2 \sumk \frac{S_k}{R_k\eik^2}
+ 4 S_i \sumk \frac{1}{R_k\eik^2} \Bigg]
-2\pi i \tauo + {\rm const}
\eea
whose solution, to $\cO(\La^{4N})$, is 
\bea
\label{Svev}
\vev{S_i} &=& 
{\alpha \over R_i}  \La^{2N} 
+ {\alpha \over R_i}  \La^{4N} 
\Bigg[
\muteplus {3\over 2 R_i^2} \sumk \suml {1\over \eik\eil}
+ 4 \sumk \suml \frac{1}{R_k R_\ell \eik \ekl} \non\\
&& 
+ \frac{2}{R_i^2} \sumk \frac{1}{\eik^2} 
- \frac{2}{R_i} \sumk \frac{1}{R_k \eik^2}
+ 2 \sumk \frac{1}{R_k^2\eik^2}
\Bigg] + \cO(\La^{6N})
\eea
where $\tauo$ and the other constants 
in eq.~\eqs{extreme} have been absorbed
into a redefinition of the cut-off
$\La = {\rm const}\, {\times} \hLa \, \e^{\pi i \tauo/N}$.
(This definition of $\La$ corresponds to that used
in the Seiberg-Witten curve \eqs{Sigma}.)

Although we are primarily interested in the $\cN=2$ limit in
this paper, 
the $\cN=1$ effective superpotential may be easily 
computed by substituting eq.~\eqs{Svev} into eq.~\eqs{effective}.

We can now evaluate 
\beq
\tau_{ij}(e)  = {1\over 2\pi i} 
{\partial^2 \free \over \partial S_i \partial S_j}
\bigg|_{S_i = \vev{S_i} } 
= \tau_{ij}^{\rm pert}(e) + \sum_{d=1}^\infty \La^{2Nd} \tau_{ij}^{(d)}(e)
\eeq
to obtain the perturbative contribution
\beq
\label{taupert_of_e}
2\pi i \tau^{\rm pert}_{ij} (e)
=\delta_{ij} 
\Bigg[ {\rm const} - \sumk \log \left( \eik \over \La \right)^2 \Bigg]
+ (1 - \delta_{ij}) 
\Bigg[ {\rm const} + \log \left( \eij \over \La \right)^2 \Bigg]
\eeq
and the one-instanton contribution
\bea 
\label{tau1_of_e}
&&2 \pi i \tau^{(1)}_{ij} (e)= 
\de_{ij} \left[ 
 {8 \over R_i^2} \sumk \suml {1\over \eik\eil}
- 4  \sumk \suml {1\over R_k^2  \eik\ekl}
+ {10\over R_i^2} \sumk {1\over \eik^2}
+ 10 \sumk {1\over R_k^2 \eik^2} \right]\non\\
&&+ (1-\de_{ij})
\left[ 
- {8 \over R_i^2} \sum_{k\neq i,j} {1\over \eij \eik}
- {8 \over R_j^2} \sum_{k\neq i,j} {1\over \eji \ejk}
+ 4  \sum_{k\neq i,j}  {1\over R_k^2  \eik \ejk}
- {10\over R_i^2\eij^2}
- {10\over R_j^2\eij^2}
\right]
\eea
to the gauge coupling matrix.
We have repeatedly used the identity
\be 
\label{id}
\sumk {1 \over R_k \eik} = - {1\over R_i} \sumk {1\over \eik}
\eeq
which can be derived by taking the $z \to e_i$ limit of both sides of
\beq
\prod_{k=1}^N {1 \over z-e_k}  
- {1 \over R_i (z-e_i)} = \sum_{k \neq i} {1 \over R_k (z-e_k)}\,.
\eeq
Finally,  we take the limit $\alpha \to 0$ to restore $\cN=2$ supersymmetry,
but this has no effect on $\tau_{ij}$, which is independent of $\alpha$.

The logarithmic terms in eq.~\eqs{taupert_of_e} 
reflect the running of the coupling constants
of this asymptotically free theory.

The gauge couplings $\tau_{ij}$ are usually written in terms of 
the periods $a_i$, which are related to the $e_i$'s by
$a_i =e_i + \cO(\La^{2N})$. 
{}From this it can be seen that one may write
the perturbative contribution to the gauge couplings as
\beq
\label{taupert_of_a}
2\pi i \tau^{\rm pert}_{ij} (a)
=\delta_{ij} 
\Bigg[ {\rm const} - \sumk \log \left( a_i - a_k \over \La \right)^2 \Bigg]
+ (1 - \delta_{ij}) 
\Bigg[ {\rm const} + \log \left( a_i - a_j \over \La \right)^2 \Bigg]
\eeq
which implies that the perturbative prepotential is
\be
2\pi i\cF_{\rm pert} (a) 
= -{\ts \frac{1}{4}} \sum_i \sum_{j\neq i} (a_i-a_j)^2 
\log \left(a_i-a_j \over {\rm const} \times \La \right)^2 
\ee
in agreement with the well-known result. 
To obtain the one-instanton contribution to the prepotential,
$\cF^{(1)}(a)$,
from perturbative matrix theory,
however, one needs to know the $\cO(\La^{2N})$ correction to the relation  
between $a_i$ and $e_i$. 
We turn to this question in the next section, 
and then return to the computation of 
$\cF^{(1)}(a)$ in section 6.

\section{Determination of $a_i$ within the matrix model}
\setcounter{equation}{0}

In Seiberg-Witten theory, 
$a_i$ is the $A_i$-period integral of $\laSW$.
How is $a_i$ defined in the context of the perturbative matrix model?

To motivative the conjecture below, we first consider 
\beq
\label{udef}
u_n = {1 \over n} \tr (\phi^n) 
\eeq
where $\phi$ is the scalar component of the adjoint 
$\cN=1$ chiral superfield of the $\cN=2$ vector multiplet.
In the Seiberg-Witten approach, the vevs of these operators
may be written in terms of integrals over the $A_i$ 
cycles \cite{Dijkgraaf:2002d}:
\beq
\label{uint} 
\vev{u_n} = {1 \over 2 \pi i n} \sumN \oint_{A_i}  x^{n-1} \laSW  \,.
\eeq
On the matrix model side,
$\vev{u_n}$ may be computed via\footnote{We 
thank Cumrun Vafa for this explanation.} 
\beq
\label{ucalc} 
\vev{u_n} = 
{ \pa \tWeff(e,\vev{\tS},\eps) \over \pa \eps } \bigg|_{\eps \to 0}
\eeq
where $\tWeff(e,S,\eps)$ is the effective superpotential that one obtains
by considering the matrix model with action
$\tW(\Phi) =  W(\Phi) + \eps (1/n)\tr (\Phi^n) $.\footnote{For 
$n \le N+1$, $\tW(\Phi)$ is equivalent to $W(\Phi)$ (\ref{W}), with 
$\alpha s_{N+1-n} \to \alpha s_{N+1-n} + \eps$,
so \eqs{ucalc} becomes \cite{Cachazo:2001}
\beq
\vev{u_n} = 
{1\over \alpha} { \pa \Weff(e,\vev{S}) \over \pa s_{N+1-n} }.
\eeq
}
Spelling this out more explicitly, 
one considers 
\beq
\tZ = \exp \left(   {1\over \gs^2} \tfree  \right)
= {1\over \mathrm{vol}(G)} 
\int  \D\Phi  \exp 
\left( \minus \, {1\over \gs} 
\left[ W(\Phi) + \eps\, {1\over n}\tr(\Phi^n)\right] \right)
\planar
\eeq
Then, writing $\tfree = \free + \eps \delF$, one computes
\beq
\label{tWeff}
\tWeff (e,S, \eps) =  \minus \sumN  N_i {\partial \tfree \over \partial S_i}
\plus 2 \pi i \tauo \sumN  S_i
=  \Weff (e,S) \minus \eps \sumN N_i {\pa \over \pa S_i} \delF \,.
\eeq
Extremizing $\tWeff (e,S, \eps)$ with respect to $S$ gives
$\vev{\tS_i} = \vev{S_i} + \eps \delS_i + \cO(\eps^2)$.
Substituting $\vev{\tS}$ into eq.~\eqs{tWeff},
one obtains
\beq
 \tWeff (e,\vev{\tS}, \eps) 
=\Weff (e,\vev{S}) + \eps \sumN \delS_i {\pa \Weff  \over \pa S_i} \vevS
\minus \eps \sumN N_i {\pa \over \pa S_i} \delF \vevS + \cO(\eps^2)
\eeq
The second term vanishes by the definition of $\vev{S}$.
Finally, using eq.~\eqs{ucalc}, one obtains
\beq
\vev{u_n} = \minus \sumN N_i {\pa \over \pa S_i} \delF \vevS
\eeq
Now observing that to first order in $\ep$
\beq
\tZ = 
{1\over \mathrm{vol}(G)} 
\int  \D\Phi  \exp 
\left( \minus \, {W(\Phi) \over \gs} \right) \planar
\! + \;
{1\over \mathrm{vol}(G)} 
\int  \D\Phi  
\left[ \minus\,  {\eps \over \gs n} \right] \tr (\Phi^n)
\exp \left( \minus \, {W(\Phi) \over \gs} \right)
\planar
\eeq
we see that $\delF$ can be obtained by computing the
(connected) planar $n$-point function 
$\vev{\tr (\Phi^n)}$ 
in the matrix model with action $W(\Phi)$,
thus giving the explicit expression
\beq
\vev{u_n} = \sum_i N_i {\pa \over \pa S_i} 
{\gs \over n} \vev{\tr( \Phi^n)}  \vevS
\eeq
In section 7, we will use this expression to compute 
the one-instanton contribution to $\vev{u_n}$.

Turning now to $a_i$, 
recall that on the gauge theory side, 
\beq
\label{aint}
a_i  = {1\over 2\pi i} \oint_{A_i} \laSW\,.
\eeq
We propose that,
just as $\vev{u_n}$ is related to $\tr(\Phi^n)$, so
$a_i$ is related to $\tr_i(\Phi)$, 
where in the latter case, 
we trace only over the $i$th diagonal block of $\Phi$. 
This prescription is motivated by the following facts. 
Whereas the contour in eq.~\eqs{uint} 
is over the sum of $A_i$ cycles, 
the contour in eq.~\eqs{aint} is 
only over a single $A_i$ cycle. 
It has been observed that 
when one sums the matrix perturbation series \cite{Dijkgraaf:2002a},
each block of eigenvalues spreads out via eigenvalue repulsion 
into a distribution along a branch cut of the spectral curve; 
thus a single block corresponds to a single branch cut. 

Considering a generic point in moduli space, 
where U$(N) \to \U(1)^N$ (so that $N_i = 1$),
we have 
\beq
a_i = \sum_j {\pa \over \pa S_j} \gs \vev{\tr_i (\Phi)}  \vevS
\eeq
Now expanding $\Phi$ around the vacuum \eqs{expand},
$\tr_i (\Phi) = M_i e_i + \tr(\Psi_{ii}) $, 
and we find
\beq
\label{amat}
a_i =  e_i + \sum_j {\pa \over \pa S_j} \gs \tadpole \vevS
\eeq
where $\tadpole$ is obtained by calculating
connected planar tadpole diagrams
with an external ${ii}$ leg  
in the matrix model.

Eq.~\eqs{amat} is our conjectured matrix model definition of $a_i$.
The right-hand side of eq.~(\ref{amat}) is independent of 
$\alpha$, and thus survives in the $\cN=2$ limit, 
as required for consistency.
One important implication of our conjecture 
is that we need only evaluate 
tadpole diagrams in the matrix model
to find the relation between $a_i$ and $e_i$. 
We stress that this procedure does not require 
knowledge of the Seiberg-Witten curve or of $\lambda_{\mathrm{SW}}$,  
and the calculation can be done order-by-order 
in the perturbative expansion.

\section{Calculation of $a_i$ and $\cF(a)$ for $\U(N)$}
\setcounter{equation}{0}

We will now test our proposal for the matrix model definition
of $a_i$ for the case of $\cN=2$ $\U(N)$ gauge theory. 
The relevant tadpole diagrams 
to first order in the instanton expansion
are displayed in figure 2.

\begin{figure}[h]
\begin{center}
 \includegraphics{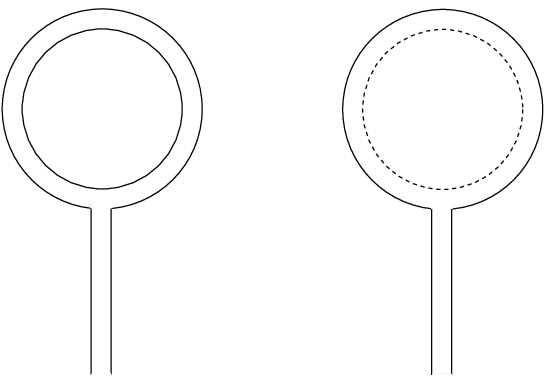}  \\[-2cm]
{\small {\bf Figure 2:} Tadpole diagrams contributing to the 
one-instanton contribution to $a_i$.}
\end{center}
\end{figure}

Using the Feynman rules derived from the action \eqs{freeenergy},
one obtains 
\be
\tadpole = 
{1 \over \al \gs} 
\sum_{j\neq i}
\left[-\frac{S_i^2 }{R_i \eij}
+ 2\frac{S_i S_j}{R_i \eij} \right]
\ee
Inserting this result into eq.~\eqs{amat},
evaluating the resulting expression using eq.~\eqs{Svev},
and using the identity \eqs{id},
we find (note that this expression is $\alpha$ independent)
\be
\label{ae}
a_i = e_i -  {2 \La^{2N}\over R_i^2}\sum_{j\neq i}\frac{1}{\eij}
+ \cO(\La^{4N})\,,
\ee
which agrees with the known result \cite{D'Hoker:1996}.
(In the appendix, we present an alternative derivation of this formula 
that uses the fact that the Seiberg-Witten differential is related
to the density of gauge theory eigenvalues in the large-$N$ 
limit \cite{Dijkgraaf:2002d}.)
Equation \eqs{ae} implies that
\beq
\log \eij  = \log \aij
+ \La^{2N} \left[
 {2 \over R_i^2} \sum_{k\neq i,j} {1\over \eij \eik}
+{2 \over R_j^2} \sum_{k\neq i,j} {1\over \eji \ejk}
+{2\over R_i^2\eij^2}
+{2\over R_j^2\eij^2} \right]
\eeq
where $\aij= a_i - a_j$.
We can now re-express $\tau_{ij}$ 
(\ref{taupert_of_e}), (\ref{tau1_of_e})
in terms of $a_i$
\beq
\label{tau_of_a}
\tau_{ij} (a)  =
\tau_{ij}^{\rm pert}(a) + \sum_{d=1}^\infty \La^{2Nd} \tau_{ij}^{(d)}(a)
\eeq
where the perturbative contribution is as found above \eqs{taupert_of_a}
\beq
2\pi i \tau^{\rm pert}_{ij} (a)
=
\delta_{ij} 
\Bigg[ {\rm const} - \sumk \log \left( \aik  \over \La \right)^2 \Bigg] 
+ (1 - \delta_{ij}) 
\Bigg[ {\rm const} + \log \left( \aij  \over \La \right)^2 \Bigg]
\eeq
and the one-instanton contribution is
\bea 
&&2 \pi i \tau^{(1)}_{ij} (a) =
\de_{ij} \left[ 
 {4 \over R_i^2} \sumk \suml {1\over \aik\ail}
+ {6\over R_i^2} \sumk {1\over \aik^2}
+ 6 \sumk {1\over R_k^2 \aik^2} \right]  \\
&&+ (1-\de_{ij})
\left[ 
- {4 \over R_i^2} \sum_{k\neq i,j} {1\over \aij \aik}
- {4 \over R_j^2} \sum_{k\neq i,j} {1\over \aji \ajk}
+ 4  \sum_{k\neq i,j}  {1\over R_k^2  \aik \ajk}
- {6\over R_i^2\aij^2}
- {6\over R_j^2\aij^2}
\right] \non
\eea
where now $R_i = \prod_{j \neq i} (a_i - a_j)$.
It is readily verified that this can be written as 
$ \tau_{ij} = {\pa^2 \pre / \pa a_i \pa a_j}$
with
\beq
2\pi i\pre
= -{\ts \frac{1}{4}}\sum_i \sum_{j\neq i} (a_i-a_j)^2 
\log \left(a_i-a_j \over {\rm const} \times \La \right)^2 
+ \La^{2N} \sum_i \prod_{j \neq i} {1\over (a_i-a_j)^2}
+ \cO(\La^{4N})
\ee
This precisely agrees 
with the result obtained in eq. (4.34) of ref.~\cite{D'Hoker:1996}.

To conclude, we have shown that a completely perturbative matrix model 
calculation, 
which does not use the Seiberg-Witten curve or differential,
gives the correct result for the prepotential to first order 
in the instanton expansion.
Higher-instanton corrections to the prepotential 
may be obtained by higher-loop contributions to the matrix 
model free energy and tadpole diagrams.

\section{Calculation of $\vev{u_n}$ in the matrix model}
\setcounter{equation}{0}

In section 5, we showed that the gauge theory 
invariant $\vev{u_n} =(1/n) \vev{ \tr(\phi^n) }$ 
can be expressed in terms of a matrix model $n$-point function as
\beq
\label{umat}
\vev{u_n} = \sum_i N_i {\pa \over \pa S_i} 
{\gs \over n} \vev{\tr( \Phi^n)}  \vevS \,.
\eeq
As a check of eq.~(\ref{umat}) we now evaluate this expression to one-instanton 
order in the $ \U(N) \to \U(1)^N$ theory (so $N_i=1$).

First one expands $\Phi$ around the vacuum \eqs{expand}, 
using $\Psi_{ij}=0$ ($i\neq j$)
\beq
\tr (\Phi^n) 
= \sumN \sum_{\ell=0}^n 
\pmatrix{ n \cr \ell \cr} 
\tr \left(e_i^{n-\ell}\Psi_{ii}^\ell \right)   =
\sumN \left[ M_i e_i^n + 
 n e_i^{n-1} \tr( \Psi_{ii} ) 
+ {n(n{-}1)\over 2} e_i^{n-2} \tr( \Psi_{ii}^2 ) + \cdots \right]
\eeq
By counting powers of $S_i$ of the diagrams, 
it is not hard to see that only the $\ell \le 2$ terms will 
contribute to the one-instanton term.
The tadpole term $\vev{\tr(\Psi_{ii})}$ was already computed 
in the previous section.
To quadratic order, the only diagram contributing to 
$\vev{\tr(\Psi_{ii}^2)}$ is a $\Psi_{ii}$ loop,
giving $ \gs M_i^2 / \al R_i$.  Thus,
\beq
{\gs \over n} \vev{ \tr (\Phi^n)} = 
{1 \over n} \sumN S_i e_i^n + 
{1\over \al}  \left[
\sumN e_i^{n-1} 
\sum_{j\neq i}
\left( -\frac{S_i^2 }{R_i \eij}
+ 2\frac{S_i S_j}{R_i \eij} \right)
+ {n-1 \over 2} e_i^{n-2} {S_i^2 \over  R_i} \right] + \cO(S^3)
\eeq
Substituting this into  eq.~\eqs{umat}, one obtains
the $\alpha$-independent expression
\beq
\vev{u_n} 
= {1 \over n} \sumN e_i^n + 
\La^{2N} \left[
  2 \sumN \sum_{j \neq i}  {e_i^{n-1}  \over R_i R_j \eij}
+ (n-1) \sumN {e_i^{n-2} \over  R_i^2} \right] + \cO(\La^{4N})
\eeq
The first term is just the classical vev of $u_n$.
Using the identity (\ref{id}) the term in square brackets can be written
\be \label{ressum}
  -2 \sumN \frac{e_i^{n-1}}{R_i^2} \sum_{j \neq i}  {1 \over \eij}
+ (n-1) \sumN {e_i^{n-2} \over  R_i^2} = \sumN \frac{\pa}{\pa e_i} 
\frac{e_i^{n-1}}{R_i^2}\,.
\ee
Now consider $z^{n-1}/\prod_{j}(z-e_j)^2$. This function has double 
poles at $z=e_i$. The sum of the residues at these poles is exactly 
equal to the sum that appear on the right hand of the equality 
in (\ref{ressum}). Thus provided that there is no residue at infinity 
the sum vanishes. This is the case for $n < 2N$;
thus, there is no one-instanton correction to $(u_n)_{\rm cl}$ 
for $n < 2N$.
This is consistent with the exact result \cite{Cachazo:2002} 
$\vev{u_n} = (u_n)_{\rm cl}$ for $n \le N+1$,
which should hold to all orders in matrix model perturbation 
theory.

For $n\ge 2N$, however, the term in square brackets does not vanish, since 
for this case the residue at infinity is equal to the residue at $w=0$ 
of $-w^{2N- n - 1}/\prod_{j}(1 - w e_j)^2$. The sum above is equal to 
minus the residue at infinity, and hence equals ($m=n-2N\ge 0$)
\be
\frac{1}{m!}\left(\frac{d^m}{dw^m}\right)
\left[\frac{1}{\prod_j(1-w e_j)^2} \right] \bigg|_{w=0}\,.
\ee
For example, it is exactly equal to 1 for $n=2N$,
yielding 
\beq 
\vev{u_{2N}} = {1 \over 2N} \sumN e_i^{2N} +  \La^{2N} + \cO(\La^{4N})\,.
\eeq
It may be readily verified by deforming the contour, 
and evaluating the residue at $x=\infty$,
that the gauge theory expression \eqs{uint}, 
which uses the Seiberg-Witten differential,
yields precisely the same result, 

Matrix model perturbation theory thus provides an alternative way
of evaluating $\vev{ \tr (\phi^n) }$ in $\cN=2$ U$(N)$ gauge theory.

\section{Conclusions}

The remarkable results of Dijkgraaf, Vafa, and collaborators indicate 
that several non-perturbative results in supersymmetric gauge 
theories can be obtained from perturbative calculations in  
auxiliary matrix models, without reference to string/M-theory. 

The Seiberg-Witten approach to $\cN=2$ gauge theories requires 
the knowledge of a Seiberg-Witten curve and one-form, 
where the most general method of obtaining these involves  
M-theory \cite{Witten:1997}. 

By contrast, in the matrix model approach to $\cN=2$ gauge theories, 
one expects that all the relevant information should be contained 
within the matrix model itself. Previously, the only way to obtain 
the periods $a_i$ has been via the Seiberg-Witten differential, which was 
obtained from the restriction of a three-form in the Calabi-Yau 
setup \cite{Cachazo:2002}. However, this approach falls outside the spirit 
of the Dijkgraaf-Vafa program, 
as all gauge theory quantities should be derivable 
without reference to string theory. One of the main results of this 
paper is to provide the missing link that allows us to compute 
the periods $a_i$ of $\cN=2$ gauge theories by means of a 
perturbative calculation in the matrix model. We have shown 
that the prescription reproduces previously known results.

To obtain explicit expressions for the periods $a_i$, $a_{Di}$, and 
the prepotential $\cF(a)$ from knowledge of the curve requires 
extensive calculations (see, e.g., ref.~\cite{D'Hoker:1996}). Our 
computations are somewhat simpler than such calculations. 
However, it should be emphasized that there are other methods for 
obtaining the $\cN=2$ instanton expansion. One is via the solution 
of Picard-Fuchs differential equations \cite{Klemm:1995}; 
this method quickly becomes cumbersome as the rank of the 
gauge group increases. 
A promising technique involves recursion relations relating multi-instanton 
results to the one-instanton results \cite{Chan:1999}. 
Other methods utilize the connection 
to integrable models \cite{Gorsky:1995} and Whitham 
hierarchies \cite{Edelstein:1998}. 
It would be interesting 
to see how the above strategies manifest themselves in the matrix model setup. 
Most importantly, there are other methods which also do not make 
reference to string/M-theory. In this context we note the beautiful 
work of Nekrasov \cite{Nekrasov:2002}. It would be very interesting to 
connect this approach to that of the matrix model.

Each of the above methods has certain advantages for particular 
aspects of $\cN=2$ theories. The matrix model approach promises to give a 
number of new insights into the structure of $\cN=2$ gauge theories 
and their relations to string theory. In 
its present form the approach seems to be less computationally 
efficient than the state-of-the-art methods of Nekrasov \cite{Nekrasov:2002}, 
although there might be some models for which the matrix model 
approach offer certain advantages. 
We should also mention that the 
matrix model approach to $\cN=2$ theories as presented here
is rather roundabout. 
A more direct approach 
would be desirable. In ref.~\cite{Dijkgraaf:2002b}, a more direct route 
was proposed for $\SU(2)$ by relating this case to a  
double scaling limit of a unitary matrix model. It is 
not obvious to us, however,  how to extend this to more general models.

\section*{Acknowledgments}
We would like to thank C.~Vafa for valuable discussions. 
HJS would like to thank the string theory group and Physics 
Department of Harvard University for their hospitality extended 
over a long period of time.

\setcounter{equation}{0}
\def\theequation{A.\arabic{equation}}

\section*{Appendix}

In this appendix, we will derive the 
relation between $a_i$ and $e_i$ in an expansion in $\La$ 
purely within the context of Seiberg-Witten theory. 
This calculation was done in a very elegant fashion 
in ref.~\cite{D'Hoker:1996}.
We will present an alternative route to the same result,
based on the relation between the Seiberg-Witten differential
and the density of gauge theory eigenvalues in
the large-$N$ limit \cite{Dijkgraaf:2002d}.

Consider the U$(KN)$ theory on the dimension $N$ subspace 
of the Coulomb branch at which the symmetry is broken only to U$(K)^N$.
The SW curve $\tSig$ on this subspace 
takes the form \cite{Douglas:1995},\cite{Cachazo:2001}
\beq
\label{ukn}
\tSig: \hskip.2in 
y^2 = P_{KN}(x)^2 - 4 \tLa^{2KN},  \hskip.5in
P_{KN}(x) = \prodN \prodK  (x-\te_{ij}) 
= \tLa^{KN} T_K(P_N(x)/\La^N),\hskip.5in
\eeq
where
\beq
\label{chebyshev}
T_K(x)= 2 \cos\left[ K \arccos \left( \half x \right) \right]
\eeq
are the first Chebyshev polynomials.
Using 
$ T_K(x)^2 - 4 = (x^2 - 4) U_{K-1} (x)^2 $,
where $U_K(x)$ are the second Chebyshev polynomials,
it follows that
\beq
y^2 = \tLa^{2KN} \La^{-2N} [U_{K-1} (P_N(x)/\La^{N})]^2 (P_N(x)^2 - 4 \La^{2N})
\eeq
that is, $\tSig$ shares the $2N$ branch points of $\Sig$,
and the other $2N(K-1)$ branch points coalesce in pairs
to give $N(K-1)$ double zeros of $y^2$.
In other words, the $KN$ branch cuts of the U$(KN)$ Seiberg-Witten
curve at a generic point in its moduli space 
merge along this subspace of the moduli space
to give the $N$ branch cuts of the U$(N)$ theory.
Thus, the $A_i$ cycle of $\Sig$ is 
the sum $\sum_{j=1}^K \tilde A_{ij}$ of cycles of $\tSig$.

{}From eqs.~\eqs{ukn} and \eqs{chebyshev}, the roots $\te_{ij}$ of
$P_{KN}(x)$ obey
\beq
\label{eij}
P_N(\te_{ij} ) 
= 2 \La^N \cos \argu,\hskip.2in
j = 1, \ldots K
\eeq
that is 
\beq
\te_{ij} - e_i 
= { 2 \La^N  \over \prod_{k\neq i} (\te_{ij} - e_k) } \cos \argu,
 \eeq
This equation can be iteratively solved for $\te_{ij}$ giving
\beq
\label{eijsoln}
\te_{ij} = e_i 
+ { 2 \La^N  \cos \argu \over \prod_{k\neq i} (e_i - e_k) }
- { 4 \La^{2N}  \cos^2 \argu \over \prod_{k\neq i} (e_i - e_k)^2 }
\sum_{\ell\neq i} {1 \over (e_i - e_l) }
+  \cdots
\eeq

Moreover, using eq.~\eqs{eij}, one may calculate that,
in the large $K$ limit,
the density $\sigma(x)$ of $\te_{ij}$ along the
branch cut is
\beq
\sigma (x) = {K\over\pi}  {P_N^\prime(x) \over \sqrt{4 \La^{2N}-P_N(x)^2  }}
\eeq
Comparing this to eq.~\eqs{periods} we see that 
\beq
\sigma(x) \D x = {K \over \pi i x} \laSW 
\eeq
which is no accident, as we will now see. 

Again using eq.~\eqs{chebyshev}, one
may compute that the SW differential of the U($KN$) theory
along this subspace of the Coulomb branch is proportional to 
that of the U($N$) theory:
\beq
\label{laSWforKN}
\tlaSW = { x P_{KN}^\prime (x)  \D x \over \sqrt{ P_{KN}^2 (x) - 4 \tLa^{2KN}}}
=  { K x P_{N}^\prime (x)  \D x \over \sqrt{ P_{N}^2 (x) - 4 \La^{2N}}}
= K \laSW
\eeq
Thus
\beq
\label{aiaij}
a_i = {1\over 2\pi i}\oint_{A_i} \laSW 
= {1\over 2\pi i K} \sum_{j=1}^K \oint_{\tA_{ij}} \tlaSW 
= {1\over K} \sum_{j=1}^K \ta_{ij}
\eeq
Also, using eq.~\eqs{uint}, it immediately follows from
eq.~\eqs{laSWforKN} that
$\vev{\tu_n} = K \vev{u_n} $ for $ n=1,\cdots,N$,
where $\vev{\tu_n}$ are the vevs of $(1/n) \tr(\phi^n)$ in SU($KN$).
Equation \eqs{aiaij} holds for all $K$, so we take $K$ large.
In the $K \to \infty$ limit, $\ta_{ij} = \te_{ij}$, thus 
\beq
\label{aieij}
a_i = \lim_{K \to \infty} {1\over K} \sumK \te_{ij}  
\eeq
In the large $K$ limit, the sum over $j$ can be replaced with
an integral over the density of $\te_{ij}$'s along the $i$th cut
\beq
a_i =  {1\over K} \int_{i^{\rm th}~{\rm cut}}  x \sigma(x) \D x
   =  {1\over \pi i } \int_{i^{\rm th}~{\rm cut}}  \laSW
\eeq
which is simply our starting point \eqs{periods},
since the integral along the cut is exactly half an $A_i$ cycle.
Moreover, in the large $K$ limit,
we can use eq.~\eqs{eijsoln} in eq.~\eqs{aieij}
to obtain
\beq
a_i = e_i - { 2 \La^{2N}  \over \prod_{k\neq i} (e_i - e_k)^2 }
\sum_{\ell\neq i} {1 \over (e_i - e_l) }
+  \cdots
\eeq
This agrees with the results of ref.~\cite{D'Hoker:1996},
and with the matrix model calculation presented in the main body
of this paper.

\end{document}